\documentclass[11pt,titlepage,final]{amsart}
\usepackage{epsfig}
\usepackage{graphicx}
\usepackage{epsfig}
\usepackage{amssymb}
\newcommand{\diff}{{\rm d}}
\newcommand{\MM}{{\sc mm}}
\newcommand{\half}{\tfrac12}
\newcommand{\Mop}{\mathcal{M}}
\newcommand{\h}{\mathfrak{h}}
\newcommand{\Lop}{\mathcal{L}}
\newcommand{\Hop}{\mathcal{H}}
\newcommand{\Top}{\mathcal{T}}

\newcommand{\Cop}{\mathcal{C}}
\newcommand{\derx}{\partial_x}
\newcommand{\derq}{\partial_x^2}

\begin{document}
  \begin{titlepage}
    \begin{flushright}
      {\small\sf LPM/04-17}\\
      {\small\sf UPRF-2004-09}
    \end{flushright}
    \vskip 1.in
    \begin{center}
      \textbf{\large Exact and semiclassical approach to a class of singular
        integral operators arising in fluid mechanics and quantum field theory}\\[3.em]
      \textbf{\large V.\ A.\ Fateev\footnote{\it Laboratoire de Physique
          Math\'{e}matique, Universit\'{e} Montpellier II, Pl. E.\ Bataillon,
          34095 Montpellier, France, {\sf
            fateev@lpm.univ-montp2.fr}}$^{,}$\footnote{On leave of absence from
          Landau Institute for Theoretical Physics, ul.Kosygina 2, 117940
          Moscow, Russia.}}, \textbf{\large R.\ De Pietri$^3$}\hspace{0.1cm}
      {\large and} \textbf{\large E.\ Onofri\footnote{\it Dipartimento di
          Fisica, Universit\`a di Parma, and {\small\sf I.N.F.N.}, Gruppo
          Collegato di Parma, 43100 Parma, Italy, {\sf
            \mbox{{$\scriptstyle [name]$}@unipr.it }}} }\\[2.em]
    \end{center}
    
  \bigskip
  \bigskip
  \begin{center}
    \textbf{Abstract}
  \end{center}
  {\small A class of singular integral operators, encompassing two physically
    relevant cases arising in perturbative QCD and in classical fluid dynamics,
    is presented and analyzed. It is shown that three special values of the
    parameters allow for an exact eigenfunction expansion; these can be
    associated to Riemannian symmetric spaces of rank one with positive,
    negative or vanishing curvature.  For all other cases an accurate
    semiclassical approximation is derived, based on the identification of the
    operators with a peculiar Schroedinger-like operator.  }
\end{titlepage}

\section{ Introduction }\label{sec:intro}
It has recently been realized that a special kind of singular integral equation
arising in the study of jet production ($e^+e^-\rightarrow q\,\bar q$ + anything)
\cite{Marchesini:2003nh,marchesini04:_exact_solut_bfkl} bears a striking
similarity to another equation\footnote{We are indebted to R.\ A.\ Askey for
  pointing out to us the reference to Tuck's paper.} introduced fourty years ago
by E.\ Tuck \cite{tuck64} in the context of laminar flows around slender bodies.
In this note we describe a general two-parameter family of integral operators
which reduce to Tuck's and Marchesini--Mueller's (hereafter \MM) case for
special values of the parameters:
\begin{equation*}
  \begin{split}
    (  K_{\alpha \beta}\,\phi)(x) &\equiv \int_{-1}^1\,\frac{\phi(x)-\phi(y)}{|x-y|}\,\diff y \,+\; \\
    +& \left((1-\alpha)\,\log(1+x) + (1-\beta)\,\log(1-x)\right)\,\phi(x)\;.
  \end{split}
\end{equation*}
We shall show that the spectral problem can be solved exactly in three cases,
connected to the three distinct symmetric spaces of rank one (with curvature 1,
0, -1) and they correspond to $(\alpha, \beta)=(1,1), (0,0), (0,1)$,
respectively. The first case \cite{tuck64} has discrete spectrum and it is
(unitarily equivalent to) a function of the Laplacian on the sphere restricted
to the axially symmetric sector. The second case has continuous spectrum and it
is (unitarily equivalent to) a function of the Laplacian on the real line. The
third case is equivalent to Marchesini--Mueller's operator and it is (unitarily
equivalent to) a function of the radial Laplacian on the hyperbolic plane.
Tuck's case corresponds to $\Top \equiv \,\half K_{2\,2}$ and no exact solution
is presently known.  To it we can nonetheless apply a semiclassical
approximation (WKB) which will be derived in general for any positive value of
the parameters; the spectrum is purely discrete in this case and it is
approximated by
\begin{equation}\label{eq:wkb}
  \begin{split}
   \kappa^{( \alpha,\beta)}_{n} &\approx 2\,\bigg[\log (\pi (n+\half))-\log \left(
      \frac{\Gamma(\alpha/2)\,\Gamma (\beta/2)}{\Gamma ((\alpha +\beta)/2)}\right) - \\
    & + (1-\half(\alpha +\beta))\,\log 2+\gamma _{E}\bigg]
  \end{split}
\end{equation}
($\gamma_E$ is Euler's constant), which in particular gives Tuck's eigenvalues
to a very good accuracy (see Table 1). 

\par\vskip .1in
This note is organized as follows.  In Section~\ref{sec:march-muell-equat} we
introduce the special problem related to jet-physics and we show that it is
unitarily equivalent to $K_{0\,1}$. In Section~\ref{sec:commutative-algebra} we
identify a second order differential operator $\Lop$ commuting with $K_{0\,1}$
and determine its eigenfunction expansion. Moreover, a first order differential
operator $\ell$ is shown to commute with the operator $K_{0\,0}$, and also in
this case we can obtain the spectral representation which is used in
Section~\ref{sec:schr-repr} to introduce another representation of the
operators, equivalent to Schroedinger's equation with a kinetic energy given by
$g(p)$, where $p$ is the momentum operator $-i\diff/\diff u$ and $\,g\,$ is
essentially Lipatov's function.  In this representation it is easy to derive
qualitative properties of the operator $K$ and to set up the semiclassical
approximation.  We also derive the boundary behaviour of eigenfunctions in the
general case.  In Appendix A we show that the operator $\Lop$ commuting with
$K_{0\,1}$ is indeed equivalent to the Laplace operator on the hyperbolic plane,
a fact which gives us valuable information on the eigenfunction expansion
(completeness, spectral measure). In Appendix B we give an easy proof of Tuck's
result about $K_{1\,1}$ which is essential to the developments of
Sec.~\ref{sec:commutative-algebra}.

\section{Marchesini--Mueller's equation}\label{sec:march-muell-equat}
Marchesini and Mueller \cite{Marchesini:2003nh}, introduced an equation for the
multiplicity of quark-antiquark pairs in electron-positron collisions. As a
function of energy, the multiplicity density satisfies an evolution equation
given by
\begin{equation*}
 \frac{  \partial\,u(\tau,\xi)}{\partial \tau} =
 \int_0^1\frac{d\eta}{1-\eta}\left[\frac{u(\tau,\eta\,\xi)}{\eta}\,-u(\tau,\xi)\right]
 +\int_{\xi}^1\frac{d\eta}{1-\eta}\left[u(\tau,\xi/\eta)-u(\tau,\xi)\right]
\end{equation*}
where $\tau$ is the logarithm of the energy in the center of mass and
$\xi=\half(1-\cos\theta)$, with $\theta$ the angle between the two jets emerging
from the electron-positron collision. Knowing the multiplicity at low energy,
its energy dependence can be calculated at all higher energies by QCD
perturbation theory and the result, in a special regime, gives the integral
equation above. For details see \cite{Marchesini:2003nh}.  It is formally very
similar to the so--called BFKL equation \cite{r.k.ellis98:_qcd}.  The unknown
$u(\tau,\xi)$ is defined for $\xi\in(0,1)$, and it vanishes at $\xi=0$ to ensure
convergence. The initial value problem is solved if we can find the spectral
decomposition of the operator in r.h.s.

It's a matter of simple algebra to show that actually the equation can be recast
into the form
\begin{equation*}
\frac{\partial\,\phi}{\partial \tau} = - K_{0\,1}\,\phi
\end{equation*}
by performing the following transformation
\begin{equation*}
 u(\tau,\xi) = e^{-\log2\,\tau}\,\xi \, \phi(\tau,2\,\xi\!-\!1) 
\end{equation*}

The operators $K_{1\,1}$ and $K_{0\,1}$ play a central role in the following.
Hence we introduce a special notation for them: 
\begin{eqnarray*}
\Hop &\equiv& \half K_{1\,1}\\
\Mop &\equiv& K_{0\,1}-\log2 = 2\Hop + \log\half(1+x) \,.
\end{eqnarray*}
It is known \cite{tuck64} that $\Hop$ is diagonal in the basis of Legendre
polynomials and its discrete eigenvalues are given by the the harmonic numbers
\begin{equation*}
     \Hop\,P_n = \h_n\,P_n\,,\quad  \h_n = 
  \begin{cases}
     0& \text{for $n=0$}\\
     \sum_{j=1}^{n}\frac{1}{j} & \text{for $n>0$}
   \end{cases}
\end{equation*}
(a simple proof of this result can be found in Appendix B).  One could study the
general spectral problem for $(\alpha, \beta)$ close to $(1,1)$, {\it e.g.\/} by
perturbation theory. However for arbitrary values of the parameters a different
approach is needed. It has been shown in
Ref.\cite{marchesini04:_exact_solut_bfkl} that the operator $\Mop$ has actually
a continuous spectrum, and the eigenfunctions can be identified with
hypergeometric functions. The result is obtained by an expansion starting from a
combination of phase-shifted plane waves. The expansion can be pushed to all
orders, and the resulting series is convergent to a hypergeometric function
which can be identified with Legendre functions. The spectral decomposition of
$\Mop$ is reduced to the classical \emph{Mehler-Fock} transform. We shall come
back to these facts in Appendix.  Here we want to show how this exact result can
be derived without any approximate procedure, by looking for a local
(differential) operator commuting with $\Mop$.  This will give an alternate more
rigorous proof of the solution.

\section{Commutativity with differential operators and  exact solution} \label{sec:commutative-algebra}
\subsection{The \MM\ operator.}
The easiest way to solve \MM\ equation is to find a differential operator $\Lop$
that commutes with $\Mop$:
\begin{equation*}
[\Lop,\Mop]=[\Lop,2\Hop]+[\Lop,\log(1+x)] = 0 
\end{equation*}
It is convenient to look for the operator $\Lop$ such that $\Lop$ as well as $
[\Lop,\log(1+x)]$ act in a simple way on $P_{n}(x)$; taking into
account the known properties of Legendre polynomials
\begin{eqnarray*}
\Lop_{0}\,P_{n}&\equiv&[(1-x^2)\derq - 2 x\,\derx]\,P_n(x) = -n(n+1)P_n(x)\nonumber\\
x\,P_n &=& \frac{n+1}{2n+1}P_{n+1}+\frac{n}{2n+1}P_{n-1}\\
-(1-x^{2})\derx\, P_{n}&=&\frac{n(n+1)}{2n+1}\,(P_{n+1}-P_{n-1})\,,
\end{eqnarray*}
the problem will reduce to a purely algebraic one. Since the \MM\ equation has a
singularity at $x=-1$ we will search $\Lop$ in the form:
\begin{equation*}
\Lop = (1+x)\,\Lop_{0}+a(1-x^2)\,\derx + b(1+x)  
\end{equation*}
By construction, this operator acts in a simple way on $P_{n}(x)$.  Namely,
\begin{equation*}
\Lop\, P_n(x) = A_n\,P_{n+1}+ B_{n}\,P_n + C_{n-1}\,P_{n-1}
\end{equation*}
where\footnote{Our convention for the indices is the natural one if we think to
  the r.h.s. as the action on the left by a tridiagonal matrix with vectors
  $[C,B,A]$ along the diagonal.}
\begin{equation}\label{eq:ABC}
  \begin{cases}
  (2n+1)\, A_n = -n(n+1)^2 - a\,n(n+1) + b\,(n+1)\\
  (2n+1)\, C_{n-1} = -n^2(n+1) + a\,n(n+1) + b\,n 
\end{cases}
\end{equation}
(and $B_n$ does not enter in what follows).
The action of the commutator $[\Lop,\Hop]$ on $P_n$ is then given by
\begin{equation}
  \label{eq:commPn}
  [\Lop,\Hop]\,P_n = \h_n\Lop P_n - \Hop\Lop P_n = 
\frac1{n} C_{n-1}P_{n-1} - \frac1{n+1} A_n P_{n+1}
\end{equation}

The commutator $\Cop\equiv[\Lop,\log(1+x)]$ can also be easily
calculated:

\begin{equation}\label{eq:Cop}
\Cop = 2(1-x^2)\,\derx - 2x + (a-1)(1-x) 
\end{equation}

Notice that there are no diagonal terms coming from Eq.~(\ref{eq:commPn}), hence
the only diagonal contribution to the commutator comes from the last term in
$\Cop$, which immediately implies $a=1$.  Using the properties of $P_{n},$
we can write:
\begin{equation*}
\Cop P_{n} = R_n\,P_{n+1} + S_{n-1}\,P_{n-1} 
\end{equation*}%
with
\begin{equation*}
R_n = -2\frac{(n+1)^{2}}{2n+1}\;, \qquad
S_{n-1}=2\frac{n^{2}}{2n+1} 
\end{equation*}
\noindent
Now we can check the commutativity of $\Lop$ and $\Mop$:
\begin{equation*}
\frac2{n+1}\,A_n + R_n = 0;\quad
\frac2{n}\, C_{n-1}+S_{n-1}=0\,.
\end{equation*}%
\noindent
From the first equation we have
\begin{equation}\label{eq:A}
A_n =-\frac{n+1}{2} R_n = - \frac{(n+1)^{3}}{2n+1} \,;
\end{equation}
on the other hand we have from Eq.~(\ref{eq:ABC})
\begin{equation*}
A_n = -\frac{(n+1)(n^{2}+2n-b)}{2n+1} 
\end{equation*}%
which fixes $b=-1$. The equation for $C_{n-1}$ is automatically satisfied.
Notice that for $K_{\alpha\,1}$ with $\alpha>0$ one would find a coefficient
different from one in front of $A_n$ in Eq.~(\ref{eq:A}), hence no solution.
$\Mop$ is therefore the only operator in the family $K_{\alpha\beta}$ which
allows a commuting differential operator of the form $\Lop$. The other two
cases, alluded to in the Introduction, are connected to a different choice of
$\Lop$, and will be discussed later.

Now we can find the eigenfunctions for the \MM\ equation. They satisfy the
differential equation:
\begin{equation*}
\Lop\phi = \lambda \phi\,. 
\end{equation*}
We should look for the solutions satisfying $|\phi (x)|\rightarrow |1+x|^{-1/2}$
at $x\rightarrow -1$ and which are finite at $x\rightarrow 1$. It is convenient
to parametrize $\lambda =-1/2-2k^{2}$.  Then the equation for $\phi $ can be
written as:
\begin{equation*}
[ (1-x^{2})\derq + (1-3x)\derx - 1+
\frac{1+4k^{2}}{2(1+x)}]\,\phi =0 
\end{equation*}%
With the substitution:
\begin{equation*}
\phi =(1+x)^{\alpha }\psi(x);\quad \alpha =-1/2+ik 
\end{equation*}%
we get
\begin{equation*}
[ (1-x^2)\derq + \{2\alpha +1-(3+2\alpha )x\} \derx (\alpha +1)^{2}]\psi(x)=0\,. 
\end{equation*}%
Setting $x=1-2y$, we obtain:
\begin{equation*}
\lbrack y(1-y)\partial_y^{2}+\{1-(3+2\alpha )y\}\partial_y-(\alpha
+1)^{2}]\,\psi(y)=0 
\end{equation*}%
which is a hypergeometric equation with $a=b=\alpha +1=1/2+ik$ and $c=1$. The
solution which is finite at $y=0$ ($x=1$) and at $y=1$ ($x=-1$) has the form:
\begin{eqnarray*}
\psi(y) &=&F(\half+ik,\half+ik,1,y); \\
\phi (k,x) &=&(1+x)^{-\half+ik}\,F(\half+ik,\half+ik,1,\frac{1-x}{2})
\end{eqnarray*}
If we re-introduce the variable $\xi=(1+x)/2$ the result can be written in the
form:
\begin{equation}\label{eq:eigf}
\phi (k,\xi) = C\,\xi^{-1/2+ik}\,F(\half + ik,\half + ik,1,1-\xi) 
\end{equation}%
To calculate the eigenvalue $\kappa$ of MM as a function of $k$ (the dispersion
relation) we may use the fact that $\Hop$ annihilates the constant and it
symmetric. It follows
\begin{equation}\label{eq:kappa}
\kappa(k) \int \phi(k,\xi)\,\diff \xi = \int \log \xi\,\phi(k,\xi)\,\diff \xi 
\end{equation}
This integral can be calculated and it is given by a combination of
\emph{digamma} functions known as \emph{Lipatov's function}
\cite{r.k.ellis98:_qcd} 
\begin{equation}
  \label{eq:lipatov}
  \begin{split}
\kappa(k) &= \psi(\half+ik)+\psi(\half-ik)-2\, \psi(1) \\
&= -4\log2 + 14\zeta(3)\,k^2 -62\zeta(5)\,k^4 +O(k^6)\,,
\end{split}  
\end{equation}
($\psi(z)=\diff\log\Gamma(z)/\diff z$).  The evolution equation for \MM\ 
equation can now be solved by expanding $u(\tau,\xi)$ on the continuous basis
$\xi\,\phi(k,\xi)$. The spectral measure which defines the eigenfunction
expansion can be taken by \cite{koornwinder84:_jacob_lie} (see also Appendix).

\subsection{The case $\alpha\!\!=\!\!\beta\!\!=\!\!0$.} 

We note that another operator, namely: $2\mathcal{H}+\log (1-x^2)= K_{0\,0}$
also commutes with a differential operator. In this case it is easy to prove
that it commutes with a first order differential operator\footnote{It will be
  noticed that this operator is a multiple of the commutator $\Cop$ of
  Eq.(\ref{eq:Cop}).}
\begin{equation}
\ell = i[-(1-x^{2})\,\derx + x]= -i\sqrt{1-x^{2}}\;\derx\,\sqrt{1-x^{2}}.
\label{k00}
\end{equation}%
The eigenfunctions in this case have a simple form:
\begin{eqnarray}
\ell\,\phi(k,x) &=& k\, \phi(k,x)\nonumber\\
\phi(k,x)&=&(1+x)^{(ik-1)/2}\,(1-x)^{-(ik+1)/2}\;.
\label{00}
\end{eqnarray}
The eigenvalue $g(k)$ of $K_{0\,0}= 2\mathcal{H}+\log (1-x^{2})$ belonging to
these eigenfunctions can be calculated exactly in the same way as before
(see Eq.~(\ref{eq:kappa})) and it has the form:
\begin{equation*}
g(k)=\psi(\half(1+ik))+\psi(\half(1-ik))-2\, \psi(1) + 2\log2 \label{gk}
\end{equation*}%
(which again can be reduced to Lipatov's function).  This means that $K_{0\,0}$
coincides with a function of the first order differential operator, namely
\begin{equation}
2\mathcal{H}+\log (1-x^{2}) = g(\ell)  \label{gl}
\end{equation}%
In the next section we will use this representation to derive a semiclassical
approximation and for the analysis of the asymptotic behaviour of the solutions
$\phi(x)$ near the boundary points $x=\pm 1.$

\section{The Schroedinger representation and semiclassical analysis}\label{sec:schr-repr}
\subsection{The semiclassical spectrum.}
We can use the representation (\ref{gl}) for the operator $2\mathcal{H}+\log
(1-x^{2})$ in terms of the first order differential operator to transform
our integral equation in the form of a Schroedinger equation (with unusual
kinetic term) which is convenient for the semiclassical analysis. Consider
the equation:
\begin{equation}
K_{\alpha \beta }\;\phi =\{2\mathcal{H}+[(1\!-\!\alpha )\,\log(1\!+x)+(1\!-\!\beta)\,
\log(1\!-x)]\}\,\phi =\kappa\, \phi\; .  \label{e1}
\end{equation}
This equation can be rewritten in terms of Schroedinger equation. Namely if
we do the substitutions: $x=\tanh u$ and $\phi =\cosh (u)\Psi(u)$ the last
equation can be rewritten as:
\begin{equation}
g(-i\partial_u)\,\Psi(u)-[\alpha \log (1+\tanh u)+\beta \log (1-\tanh
u)]\Psi =\kappa \Psi .  \label{e2}
\end{equation}%
In the free case $\,\alpha\!=\!\beta\!=\!0\,$ we have the plane waves solutions
corresponding to the functions (\ref{00}) after this substitution. In the case
$\,\alpha\!=\!\beta\!=\!1\,$ we have $\Psi _{1\,1}=P_n(\tanh u)/\cosh u\,$; this
means that, by identifying $\tanh u\equiv \cos\vartheta$, and modulo a
similarity transformation, this case is related to the Laplace operator on the
two-dimensional sphere. Finally, the case $\,\alpha\!=\!0,\, \beta\!=\!1\,$
corresponds to MM equation, where also the exact solution is known.

It is convenient to slightly modify the function $g(z)$ and the eigenvalue
$\kappa $ by adding a constant shift. Let us introduce $G(z)$ and $\kappa
^{\prime }$ by:
\begin{equation}
G(z)=g(z)+2\,\psi(1);\quad \kappa ^{\prime }=\kappa +2\,\psi(1)  \label{e3}
\end{equation}%
where $\psi(1)=-\gamma _{E}.$ Then Eq.(\ref{e2}) can be rewritten as:
\begin{equation}
\left(G(p)+V(u)\right)\Psi(u)=\kappa ^{\prime }\Psi(u)  \label{e4}
\end{equation}%
where $p=$ $-i\partial_u$ and $V=-\alpha \log (1+\tanh u)-\beta \log
(1-\tanh u).$

We shall need the following asymptotic behaviour of $G(p)$:
\begin{equation}
G(p) = 
\begin{cases}
G(0)+7/2\,\zeta (3)p^{2}+O(p^{4})\,,&(p\sim 0)\\  
\log p^{2}+O(1/p^{2}),\quad (p\rightarrow\infty )  \label{e5}
\end{cases}
\end{equation}
The operator $G(p)$ takes on the role of the kinetic energy and is equivalent to
the usual operator of non-relativistic quantum mechanics in the low energy
limit. It follows that the operator $G(p)+V(u)$ has a discrete spectrum for
$\beta \geqslant \alpha >0$ (by symmetry we can restrict to the sector $\beta
\geqslant \alpha $ with no loss of generality), while it has continuum spectrum
for $\beta \geqslant \alpha =0.$ Let us notice that in this picture Tuck's
operator $\mathcal{T}$ ($\alpha = \beta =2$) is qualitatively very similar to
the ``trivial'' case ($\alpha = \beta =1$) which corresponds to operator
$\mathcal{H}$
 
 The ``Schroedinger'' representation can be used in different ways. In
 particular, we can exploit this representation to compute the semiclassical
 approximation to the eigenvalues and eigenfunctions. We note that near the
 turning points, where the kinetic term is small we can apply the standard WKB
 approach. This leads to the following (Bohr-Sommerfeld) semiclassical
 approximation to the eigenvalues:
\begin{equation}
\int_{a}^{b}\,G^{-1}(\kappa ^{\prime }-V)\,\diff u = \pi (n+1/2)  \label{BS}
\end{equation}%
where $G^{-1}$ is the inverse function of $G$ and $\,a,b\,$ are the turning
points.  It follows from Eq.(\ref{e5}) that the inverse function is
approximately $G^{-1}(\xi)=\exp(\xi/2)+O(\exp(-\xi/2)).$ In the region where the
semiclassical approximation works we can neglect all asymptotic terms besides
the first one.  In the main approximation we can also put $a\!=\!-\infty,\,
b\!=\!+\infty .$ Then equation (\ref{BS}) can be rewritten as:
\begin{equation}
\int\limits_{-\infty }^{+\infty }\,\exp (\half(\kappa ^{\prime }-V))\,\diff u=\pi (n+1/2)
\label{int}
\end{equation}%
This integral can be easily calculated and we derive for the spectrum of the
operator $K_{\alpha\beta }$:
\begin{equation*}
\kappa^{( \alpha,\beta)}_{n}\approx 2\,\left[\log (\pi (n+1/2))-\log \left( B(\alpha
/2,\beta /2)\right) + (1-\half(\alpha +\beta))\log 2+\gamma _{E}\right]\,,
\end{equation*}
where $B$ is Euler's ``beta'' function {\it i.e.\/} we arrive at
Eq.~(\ref{eq:wkb}).  This equation gives the standard approximation for the
harmonic numbers $ \h_{n}$ (the eigenvalues of $\half K_{1\,1}$)
up to $O(1/n^2)$, 
while for the eigenvalues of Tuck's operator $\mathcal{T}=\half K_{2\,2}$ it gives:
\begin{equation}
\half\kappa_n^{(2,2)}\approx \log (\pi (n+1/2))-\log 2+\gamma _{E}.
\label{TT}
\end{equation}%
The semiclassical formulas are contrasted to the numerical or exact
eigenvalues in Tab.1 for operators $\mathcal{T}$ and $\mathcal{H}$
respectively.

\begin{center}
  \begin{table}[ht]
    \begin{tabular}{|c|l|l|l|l|}\hline
      $n$ &  Ref. \cite{tuck02:_longit}& WKB & $\h_n$\hfill& \hfill WKB\hfill \\\hline
      0  &     0.2332 & 0.3357 &      0  &   -0.116 \\ 
      1  &     1.4437 & 1.4343 &  1&    0.9827 \\
      2  &     1.9409 & 1.9451 &  1.5&    1.4935 \\
      3  &     2.2833 & 2.2816 &  1.8333&    1.8300 \\
      4  &     2.5317 & 2.5329 &  2.0833&    2.0813 \\
      5  &     2.7342 & 2.7335 &  2.2833&    2.2820 \\
      6  &     2.9000 & 2.9006 &  2.4500&    2.4490 \\
      7  &     3.0440 & 3.0437 &  2.5929&    2.5921 \\
      8  &     3.1686 & 3.1689 &  2.7179&    2.7173 \\
      9  &     3.2803 & 3.2801 &  2.8290&    2.8285 \\\hline
    \end{tabular}\vskip .1in
    \caption{ The WKB spectrum of $\Top$ and $\Hop$.}
  \end{table}
\end{center}

The semiclassical eigenfunctions $\Psi ^{(sc)}(u)$ can be written in the
form:
\begin{equation*}
\Psi _{n}^{(sc)}(u)=A\,\sin \left(\int\limits_{-\infty }^{\;\;\; u}\,\exp \{\half(\kappa _{n}^{\prime }-V)\}\,\diff u\,+\pi/4 \right) \exp (-V(u)/4)  
\end{equation*}
where $A$ is a normalization factor and $\kappa _{n}^{\prime }=
\kappa_{n}-2\gamma_E.$ The integral gives the incomplete Beta function which
reduces to elementary trascendentals in the special cases cases $\,\alpha\!
=\!\beta =\!1\,$ and $\,\alpha\! =\!\beta\! =\!2\,$ corresponding to the
operators $\mathcal{H}$ and $\mathcal{T}$ \ respectively.  In the first case we
have:

\begin{equation}
\Psi _{n}^{(sc)}(u)=A_1\,\frac{\sin [(2n+1)\tan ^{-1}(e^{u})+\pi /4]}{\sqrt{\cosh
u}}  \label{LG}
\end{equation}%
with $A_1=(\half\pi+\tfrac1{2n+1})^{-1/2}$. In the case corresponding to
Tuck's operator $\mathcal{T}$ we obtain:
\begin{equation}
\Psi _{n}^{(sc)}(u)=A_2\frac{\sin [\pi /2(n+1/2)(\tanh u+1)+\pi /4]}{\cosh u}
\label{TU}
\end{equation}%
with $A_2=(1+\tfrac{2/\pi}{2n+1})^{-1/2}$.  If we rewrite these functions in
terms of original variables $x=\tanh u$ and functions $\phi (x)=\cosh (u)\Psi
(u)$ then the expression (\ref{LG}) gives the well known large $n$ asymptotics
of Legendre polynomials:
\begin{equation*}
\phi_n(\cos \theta)\sim\frac{\sin [(n+1/2)\theta +\pi /4]}{\sqrt{\sin\theta}}\,.
\end{equation*}
For  Tuck's case $(\alpha\! =\!\beta\! =\!2)$ the
semiclassical wave functions have a very simple form:
\begin{equation}
\phi _{n}(x)=A\sin [\pi /2(n+1/2)(x+1)+\pi /4]  \label{eq:SF}
\end{equation}
and they give a rather accurate description (Fig.1) of the true eigenfunctions
which can be easily computed numerically ({\it i.e.\/} by using the spectral representation
for the operator $\Hop$ on the Legendre basis).
\begin{figure}[ht] 
  \begin{center}
    \mbox{\epsfig{file=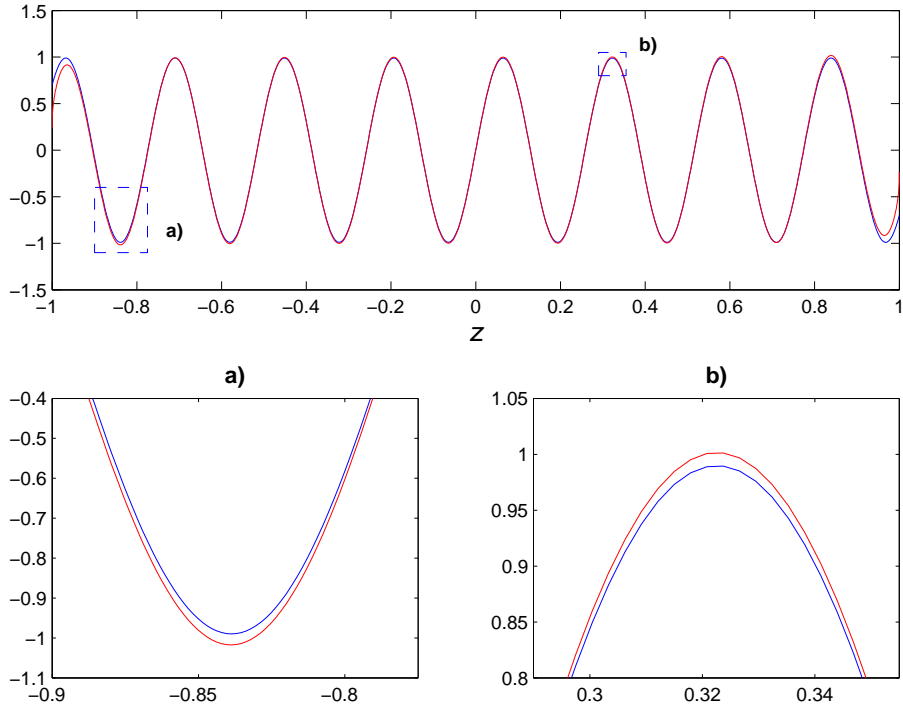,width=12.cm}}
    \caption{An example of semiclassical wave-function (Eq.~(\ref{eq:SF})) for the $\Top$ operator, $n=15$.}
  \end{center}
\end{figure}

\subsection{Boundary behaviour.}
The ``Schroedinger'' representation (\ref{e4}) can be used to derive the
asymptotic behaviour of the eigenfunctions $\phi(x)$ at the singular points
$x\rightarrow \pm 1.$ Namely we show that these asymptotics have the form:
\begin{equation}
  \phi _{\alpha \beta }\sim |\log (1+x)|^{\,d_{\alpha}}\,|\log (1-x)|^{\,d_{\beta
}}  \label{eee}
\end{equation}%
where 
\begin{equation}
d_{\alpha }=\frac{1}{\alpha }-1\,;\;d_{\beta }=\frac{1}{\beta }-1.  \label{ed}
\end{equation}

We note that the asymptotic behaviour of the function $\phi (x)$ at
$x\rightarrow \pm 1$ can be derived from the asymptotics of the function
$\Psi(u)$ at $u\rightarrow \pm \infty .$ Consider, for example, this asymptotics
at $u\rightarrow +\infty$ $(x\rightarrow 1).$ In this limit the potential term
$V$ has the form:
\begin{equation*}
V=2\beta u-2\alpha \log 2+O(e^{-2u})  
\end{equation*}
We can neglect all terms in this expansion besides the linear one. In this way
the problem is reduced to the calculation of the asymptotics at $
u\rightarrow\infty $ of the solution of Eq.(\ref{e4}) with a linear potential.
For this purpose it is convenient to rewrite this equation in the Fourier
representation.  It follows from the explicit form of function $G(p)$ that the
Fourier transform $\widetilde{\Psi }(p)$ of $\Psi(u)$ satisfies the first order
differential equation:
\begin{equation*}
-2\beta\, i\,\partial_{p}\widetilde{\Psi }(p)+[\psi(1/2+ip/2)+\psi
(1/2-ip/2)+\log 4]\widetilde{\Psi }(p)=\kappa ^{\prime }\widetilde{\Psi }(p)
\end{equation*}
The solution of this equation has the form:
\begin{equation}
\widetilde{\Psi }(p)=\left( 2^{-ip}\,\frac{\Gamma (\half(1-ip))}{\Gamma
(\half(1+ip))}\right)^{\!\!1/\beta}\;\exp (i\kappa ^{\prime }p/2\beta )  \label{FTR}
\end{equation}%
The asymptotics of the function 
\begin{equation}
\Psi(u)=\int \exp (-ipu)\widetilde{\Psi }(p)dp  \label{IFT}
\end{equation}%
is determined by the nearest singularity of the function $\widetilde{\Psi}(p)$
in the lower half plane at the point $p=-i.$ For non-integer $1/\beta $ this
singularity is a branching point. The standard estimation of the corresponding
contribution gives:
\begin{equation*}
\Psi(u)\rightarrow u^{d_{\beta }}e^{-u}(1+O(1/u));\quad u\rightarrow \infty
\end{equation*}%
Taking into account that $x=\tanh u$ and $\phi (x)=\cosh (u)\Psi(u)$ as well as
the $\,\alpha,\beta\, $ symmetry of the equation we arrive to Eqs.(\ref{eee},
\ref{ed}). It follows from Eq.(\ref{FTR}) that this asymptotics takes place in
rather narrow region $u\gg \kappa ^{\prime }/2\beta $ or $|\log (1-x)|\gg \kappa
^{\prime }/\beta $ (for $x\rightarrow 1$).

We note that for $\,\beta\! =\!1\,$ the integral (\ref{IFT}) can be calculated
explicitly and we can derive the exact wave functions in the potential
$V\!=\!2u.  $ They have the form:
\begin{equation*}
\Psi(u)=y\,J_{0}(2y)  
\end{equation*}
where $y=\exp (-u+\kappa ^{\prime }/2)$ and $J_{0}(z)$ is the Bessel function.

\section*{Acknowledgments}
{\small We thank warmly R.\ Askey for introducing us to the original paper by
  E.\ Tuck, G.\ E.\ Andrews for useful correspondence and G.\ Marchesini for
  continuous inspiration and encouragement. One of us (E.\ O.) would like to
  thank B.\ K.\ Alpert for generously providing his code for the Chebyshev
  transform, which was very useful in an early stage of this work (the
  Alpert-Rokhlin transform). E.\ O.\ would like to thank G.\ Altarelli, chairman
  of the CERN Theory Division, for his kind hospitality while this work was
  started. This work was supported by Research Training Network grant of
  European Commission contract EUCLID number HPRN-CT-2002-00325, by grant
  INTAS-OPEN-02-51-3350 and by the Italian MIUR-COFIN2003 project.}
 
\section*{Appendix A: The connection to hyperbolic plane}
The eigenfunctions $\phi(k,x)$ given by Eq.~(\ref{eq:eigf}) are related to
Legendre functions with complex index. To see this, let us revert to the
original form of \MM\ equation for which the eigenfunctions are given by
\begin{equation*}
  u(k,x) = x^{1/2+ik}\,F\left(\half + ik,\half + ik,1,1-x\right)\,.
\end{equation*}
By using well-known properties of Jacobi functions (see
\cite{koornwinder84:_jacob_lie}) we find that $u(k,x)$ can be identified with
the Jacobi function $\phi_{2k}^{(0,0)}(t)$ with $x=1/\cosh^2(t)$. Hence it
follows that $u(k,x)=P_{-1/2+ik}(2/x-1)$.  Now, it is well known that Legendre
functions of this kind appear as spherical functions on the hyperbolic
plane\footnote{see {\it e.g.\/}
  \cite{helgason84:_group_geomet_analy,gangolli72:_spher_lie}.}, {\it i.e.\/}
they are eigenfunctions of the radial part of the Laplace operator in the case
of the two-dimensional homogeneous Riemannian space with constant negative
curvature.  In Gaussian coordinates, $\diff s^2 = \diff r^2 + \sinh^2 r\,
\diff\varphi^2$, the radial part of the Laplacian $\Delta_r$ is given by
\begin{equation*}
 \Delta_r = (\diff/\diff r)^2 + \coth r\, \diff/\diff r\,.
\end{equation*}
and it is immediate to check that $(\Delta_r+(1/4\!+\!k^2))\,P_{-1/2+ik}(\cosh
r)=0$.  It is then natural to conclude that there must exist a map $x\to r$ and
a suitable similarity transformation which connects the operator $\Lop$ of
Sec.~\ref{sec:commutative-algebra} to $\Delta_r$. From the expression of
$u(k,x)$ in terms of Legendre functions, the map is given by $\cosh r = 2/x\!
-\! 1 = 4/(z+1)-1$, where $z$ is the variable which enters the definition of
$\Lop$. The similarity transformation is simply
\begin{equation*}
  \tfrac12\Lop  \equiv (1\!+\!\cosh r)\, \Delta_r \, (1\!+\!\cosh r)^{-1}\,.
\end{equation*}

Having established this connection, the explicit eigenfunction expansion comes
for free, in terms of Mehler-Fock transform
\begin{equation*}
  \begin{cases}
    u(x) = \int\limits_0^\infty\, P_{-1/2+ik}\left(\frac2{x}-1\right)\, c(k)\; \diff k\;, & (0<\!x<\!1)\\
    c(k) = k\,\tanh \pi k\,\int\limits_1^\infty\,
    u\left(\frac2{1+t}\right)\,P_{-1/2+ik}(t)\;\diff t & \end{cases}
\end{equation*}
which can be used to solve the evolution in $\tau$ for the \MM\ equation
\cite{marchesini04:_exact_solut_bfkl}.

\section*{Appendix B}
We give a simple proof of an old result due to E.~Tuck \cite{tuck64}.

\smallskip
\noindent
{\sf Theorem}. $\half K_{1\,1}$ has the Legendre polynomials $P_n(x)$ as
eigenvectors with eigenvalues the {\sl harmonic sums} $\h_n$.
\begin{proof}
By computing $K_{1\,1} \,p_n$ with $p_n(x)\equiv x^n$ we find
\begin{eqnarray*}
(K_{1\,1}\,p_n)(x) &=&  \int_{-1}^1 \diff y \,\frac{x^n - y^n}{|x-y|}\\
&=&   \left(\int_{-1}^x-\int_x^1\right)\,\diff y\,
\sum_{k=1}^n\,y^{k-1}x^{n-k}\\
&=& 2\,\h_n\,x^n - \sum_{k=1}^n\frac{1+(-1)^k}{k}\,x^{n-k}
\end{eqnarray*}
hence $K_{1\,1}$ leaves each subspace $\mathcal P_n$ of polynomials of degree
$n$ invariant for any $n$. It's matrix representation is {\sl upper triangular}
and its eigenvalues are found on the diagonal by inspection. Since $K_{1\,1}$ is
symmetric with respect to the inner product $\langle p_1,p_2\rangle =
\int_{-1}^1 dx\,p_1(x)p_2(x)$ its eigenvectors are orthogonal, hence they are
the Legendre polynomials.
\end{proof}

\bibliographystyle{unsrt} 
\bibliography{dfo}

\end{document}